\documentclass[%
amssymb,
 reprint,
nobalancelastpage,
superscriptaddress,
 amssymb,
 aps,
pra,
]{revtex4-2}
\usepackage{color}
\usepackage{graphicx}
\usepackage{dcolumn}
\usepackage{bm}

\begin{document}

\preprint{APS/123-QED}

\title{Experimental investigation of wave-particle duality relations \\ in asymmetric beam interference}

\author{Dong-Xu Chen}
 \email{xdc.81@stu.xjtu.edu.cn}
 \affiliation{Quantum Information Research Center, Shangrao Normal University, Shangrao, Jiangxi 334001, China}
\author{Yu Zhang}%
\affiliation{Quantum Information Research Center, Shangrao Normal University, Shangrao, Jiangxi 334001, China}
\affiliation{School of Physics, Nanjing University, Nanjing, Jiangsu 210093, China}
\author{Jun-Long Zhao}%
\affiliation{Quantum Information Research Center, Shangrao Normal University, Shangrao, Jiangxi 334001, China}
\author{Qi-Cheng Wu}%
\affiliation{Quantum Information Research Center, Shangrao Normal University, Shangrao, Jiangxi 334001, China}
\author{Yu-Liang Fang}%
\affiliation{Quantum Information Research Center, Shangrao Normal University, Shangrao, Jiangxi 334001, China}
\author{Chui-Ping Yang}%
 \email{yangcp@hznu.edu.cn}
 \affiliation{Quantum Information Research Center, Shangrao Normal University, Shangrao, Jiangxi 334001, China}
 \affiliation{Department of Physics, Hangzhou Normal University, Hangzhou, Zhejiang 311121, China}
\author{Franco Nori}%
 \email{fnori@riken.jp}
 \affiliation{Theoretical Quantum Physics Laboratory, RIKEN, Wako-shi, Saitama 351-0198, Japan}
  \affiliation{RIKEN Center for Quantum Computing (RQC), Wako-shi, Saitama 351-0198, Japan}
   \affiliation{Physics Department, The University of Michigan, Ann Arbor, Michigan 48109-1040, USA}

\date{\today}

\begin{abstract}
Wave-particle duality relations are fundamental for quantum physics. Previous experimental studies of duality relations mainly focus on the quadratic relation $D^2+V^2\leq1$, based on symmetric beam interference, while a linear form of the duality relation, predicated earlier theoretically, has never been experimentally tested. In addition, the difference between the quadratic form and the linear form has not been explored yet. In this work, with a designed asymmetric beam interference and by utilizing the polarization degree of freedom of the photon as a which-way detector, we experimentally confirm both forms of the duality relations. The results show that more path information is obtained in the quadratic case. Our findings reveal the difference between the two duality relations and have fundamental implications in better understanding these important duality relations.
\end{abstract}
\maketitle

\section{Introduction}
Bohr's complementarity principle initially conceptualized the controversial nature of light. It states that a photon possesses two mutually exclusive properties, wave and particle, i.e., the wave-particle duality. It is well known that this duality can be formulated by duality relations, which  quantitatively describe the trade-off between wave and particle behaviours, i.e., the emergence of one behaviour will suppress the appearance of the other. In the past years, the duality relations have drawn increasing attention because they are fundamental in quantum physics. Experimental and theoretical interests in duality relations have never vanished since the early days of quantum theory \cite{10.2307/24942949, englert1995complementarity, PhysRevLett.67.318, PhysRevA.44.4614, PhysRevA.43.492, kim2000delayed, scully1989quantum, scully1991quantum, englert1992quantum, scully1999quantum, agarwal1974quantum, englert2000mechanisms, Schleich2016, qin2019proposal}. 

The authors in \cite{wootters1979complementarity} quantified the wave-particle duality in a double-slit interference scenario and concluded that the simultaneous observation of wave and particle behaviours was possible. The complementarity relation ${P}^2+{V_0}^2\leq 1$, where ${P}$ is the predictability of the photon passing through the two paths and ${V_0}$ is the \textit{a priori} interference visibility, was derived in \cite{jaeger1995two} and \cite{englert1996fringe}. Also, Ref. \cite{englert1996fringe} considered the case in which a which-way detector (WWD) was involved, and obtained the inequality
\begin{eqnarray}
 {D_{\rm m}}^2+{V}^2\leq 1, 
 \label{eq1}
\end{eqnarray}
where $P$ was replaced by the distinguishability ${D_{\rm m}}$, and $V_0$ was replaced by the fringe visibility $V$ at the output. The predictability $P$ is different from the distinguishability ${D_{\rm m}}$ in that $P$ is the difference of the probabilities of the photon taking the two paths, while ${D_{\rm m}}$ is the which-way information stored in the WWD, which depends on the final states of the WWD and the way which we apply to retrieve the information. The distinguishability quantified in \cite{englert1996fringe} is the maximum likelihood for guessing the way right, which coincides with the minimum error discrimination (MED) of the WWD's states. 

Applying different strategies to distinguish the WWD's states gives different amounts of which-way information. Another strategy of retrieving which-way information from the WWD is the unambiguous quantum state discrimination (UQSD), which has been applied to study the wave-particle duality relation in recent years \cite{menon2018wave, len2018unambiguous, amico2020simulation, bera2015duality, qureshi2019interference, siddiqui2021multipath}. Reference \cite{menon2018wave} quantified the distinguishability by the upper bound of the probability of an unambiguous result and obtained the linear duality relation
\begin{eqnarray}
{D_{\rm u}}+{V}=1,
\label{eq2}
\end{eqnarray}
where $D_{\rm u}$ is the distinguishability derived from the UQSD strategy. A similar relation was obtained in multipath interference \cite{bera2015duality, qureshi2019interference, siddiqui2021multipath}.

The quadratic relation (\ref{eq1}) based on the MED strategy has been experimentally confirmed using various systems \cite{kim2000delayed, hellmuth1987delayed, baldzuhn1989wave, lawson1996delayed, jacques2007experimental, jacques2008delayed, kaiser2012entanglement, tang2012realization, manning2015wheeler}. However, the linear relation in Eq.~(\ref{eq2}), which is based on the UQSD strategy, has not been experimentally tested so far. Existing studies on the duality relation mostly focus on the case of symmetric beam interference, where the photon is equally likely to go through both paths. Nevertheless, the asymmetric case, where the beam splitters (BSs) are not balanced or the photon suffers from loss on the BSs, has not been investigated as much as the symmetric case. Over the past years, several theoretical analyses of asymmetric interference with a WWD have been presented \cite{li2012duality, jia2014influence, menon2018wave, liu2017complementarity, liu2019fringe}. 

In this work, {we experimentally realize asymmetric beam interference with a WWD to study the wave-particle duality.} The WWD is implemented by utilizing the polarization degree of freedom of the photon. The visibility $V$ is characterized by the fringe emerging after the interferometer. We quantify the distinguishability in two ways. One corresponds to the probability of obtaining an unambiguous result, i.e., by adopting the UQSD strategy to discriminate the WWD's states; while the other corresponds to the maximum likelihood for guessing the right path, i.e., by adopting the MED strategy to discriminate the WWD's states. 

In our experiment, both quadratic and linear forms of the duality relation, described by Eqs.~(1) and (2), {are confirmed with different degrees of asymmetry of the beam interference and different degrees of nonorthogonality of the WWD's states.} We show that the amounts of which-way information, gained through the two strategies, are different. Our experiment demonstrates the linear duality relation and also investigates the difference between the two duality relations.

\begin{figure}[hp!]
\centering\includegraphics[width=0.5\textwidth]{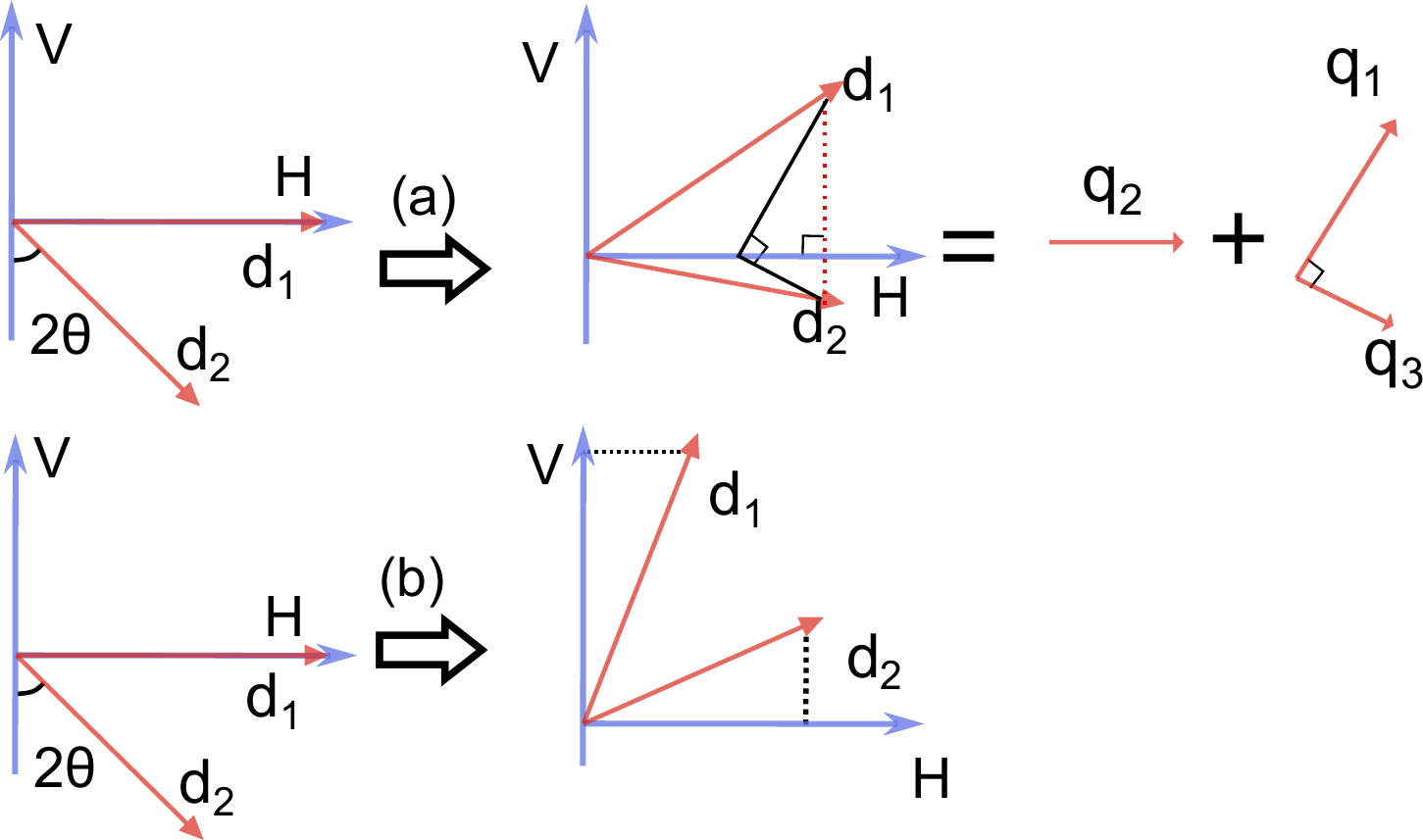}
\caption{Geometric representation of the principles of nonorthogonal state discrimination. Principles of (a) unambiguous quantum state discrimination (UQSD) and (b) minimum error discrimination (MED) strategies. The moduli of the vectors represent the square roots of the \textit{a priori} probabilities of the states. In UQSD, the states $|d_1\rangle$ and $|d_2\rangle$ are first rotated, then the horizontal component is separated into two parts, one is the common state ($|q_2\rangle$) corresponding to an inconclusive result. The residuals, $|q_1\rangle$ and $|q_3\rangle$, are orthogonal, which can be discriminated by a positive operator-valued measure. In MED, the states $|d_1\rangle$ and $|d_2\rangle$ are rotated, then a positive operator-valued measure is performed to project the states onto the basis states $|h\rangle$ and $|v\rangle$. When we detect the photon in the $|h\rangle$ state, we guess the state is $|d_2\rangle$, otherwise we guess it as $|d_1\rangle$. Since the measurement result is probabilistic, there is a probability of guessing wrongly.}
\label{fig2}
\end{figure}

\section{Results}
\subsection{Theory}
In quantum physics, quantum states, which are orthogonal to each other, can be discriminated via a single measurement. While for nonorthogonal quantum states, one is unable to discriminate them within a single measurement.  To retrieve information from nonorthogonal quantum states, different strategies are applied with different objectives. The UQSD \cite{ivanovic1987differentiate, peres1988differentiate, dieks1988overlap, mohseni2004optical, agnew2014discriminating} and the MED \cite{waldherr2012distinguishing, solis2017experimental} are the two most investigated strategies. 

Assume that one is told to discriminate two nonorthogonal states $|d_1\rangle=|h\rangle$ and $|d_2\rangle=\sin2\theta|h\rangle-\cos2\theta|v\rangle$ of a photon, with \textit{a priori} probabilities $p_1$ and $p_2$, respectively. Here, the state $|h\rangle$ ($|v\rangle$) denotes the horizontal (vertical) polarization state, and the probabilities satisfy $p_1+p_2=1$. Without loss of generality, we assume $p_2\leq p_1$. The UQSD and the MED follow different procedures as follows.

In UQSD, an unambiguous result is possible by allowing an inconclusive result. The polarization degree of freedom of the photon is coupled with another degree of freedom to form a higher dimensional space. Then $|d_1\rangle$ and $|d_2\rangle$ are projected onto the orthogonal basis
\begin{eqnarray}
|d_1\rangle=\alpha|q_1\rangle+\beta|q_2\rangle,\quad|d_2\rangle=\gamma|q_3\rangle+\delta|q_2\rangle,
\end{eqnarray} 
where $|\langle q_i|q_j\rangle|=\delta_{ij}$, $|q_1\rangle$ and $|q_3\rangle$ correspond to unambiguous results, $|q_2\rangle$ corresponds to an inconclusive result. The geometric representation of the principle of UQSD is shown in Fig.~\ref{fig2}a. The probability of obtaining an unambiguous result is $D_{\rm u}=p_1|\alpha|^2+p_2|\gamma|^2$, and it is given by \cite{jaeger1995optimal} 
\begin{eqnarray}
\label{eq4}
D_{\rm u}&=&1-2\sqrt{p_1p_2}\sin2\theta, \quad p_2/p_1> \sin^22\theta,\\
D_{\rm u}&=&p_1(1-\sin^22\theta),\quad p_2/p_1\leq \sin^22\theta.
\label{eq5}
\end{eqnarray}

On the other hand, in MED, each measurement returns a result, and the quantum state is determined by the best guess. The protocol aims at minimizing the guessing error. The principle of MED is shown in Fig.~\ref{fig2}b. The maximum probability of correctly guessing the quantum state is given by the Helstrom bound \cite{helstrom1976quantum} 
\begin{equation}
P_{\rm r}=\frac{1}{2}\left( 1+\sqrt{1-4p_1p_2\sin^22\theta}\right) .
\label{pr}
\end{equation}

\begin{figure}[tbp!]
\centering\includegraphics[width=0.4\textwidth]{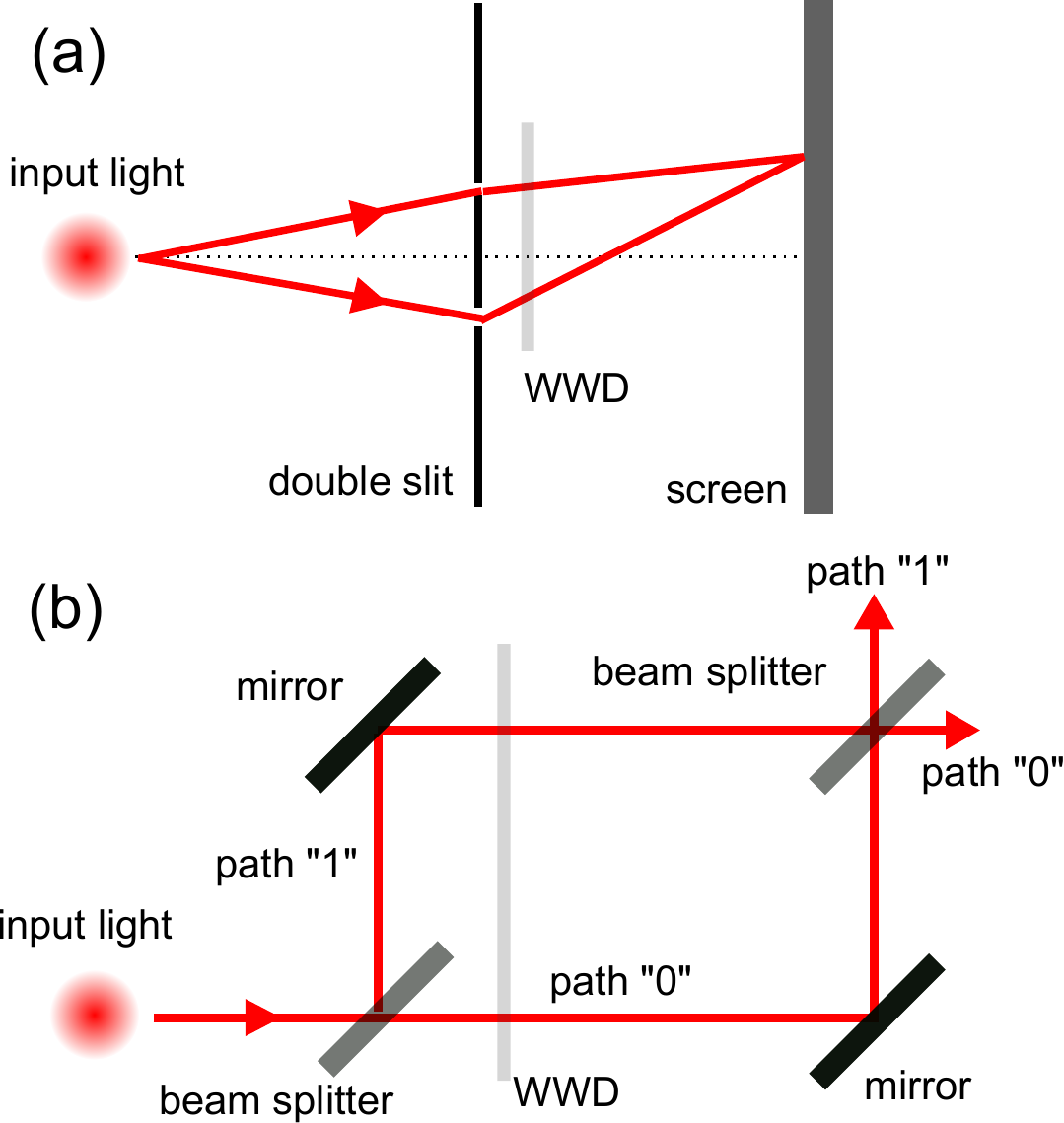}
\caption{Scenarios for study of wave-particle duality. (a) Double-slit interference setup includes a double-slit and a screen for observing the fringes. (b) A Mach-Zehnder interferometer consists of two beam splitters. WWD: which-way detector.}
\label{fig1}
\end{figure}

Research on the wave-particle duality is generally based on two scenarios, the double-slit interference and the standard Mach-Zehnder interferometer (MZI) (see Fig. \ref{fig1}). In the double-slit interference, the light passes through two separated slits, behind which a screen is placed. One can observe fringes on the screen, which are not simply the sum of the light passing through an individual slit when the other is blocked. In the standard MZI, the input light is separated by a symmetric beam splitter (BS), then the light travelling along the two arms interferes on the second BS. Changing the phase of the light in one arm gives rise to a change of the intensity of the output light. When the second BS is removed, the phase change will not induce the change of the output intensity. In this case, we declare that the photons reveal particle behaviour in an open interferometer. The double-slit interference is more often employed to intuitively show the interference of photons, while the standard MZI is more suitable in practical experiments. In studies of wave-particle duality, these two scenarios are equivalent.

The scenario we consider is an MZI with a WWD inserted in the interferometer (Fig.~\ref{fig1}b). The first BS is unbalanced, such that it causes the photon to propagate along two paths with unequal probabilities, $p_1$ and $p_2$. The WWD is a quantum detector which interacts with the photon and then gets correlated with the photon's path.

Assume now that the initial state of the WWD is $|d_0\rangle$, and the state of the photon after passing through the unbalanced BS is $\sqrt{p_1}|0\rangle+\sqrt{p_2}|1\rangle$, where $|0\rangle$ and $|1\rangle$ denote the two path states. The interaction between the WWD and the photon leads to a controlled-unitary transformation
\begin{eqnarray}
(\sqrt{p_1}|0\rangle+\sqrt{p_2}|1\rangle)|d_0\rangle\rightarrow\sqrt{p_1}|0\rangle|d_1\rangle+\sqrt{p_2}|1\rangle|d_2\rangle.
\end{eqnarray}
To distinguish the paths of the photon is equivalent to distinguish the final states of the WWD. Note that $|d_1\rangle$ and $|d_2\rangle$ are not necessarily orthogonal. When $|\langle d_1|d_2\rangle|=1$, which means $|d_1\rangle$ and $|d_2\rangle$ are identical, no path information can be retrieved from the WWD. When $|\langle d_1|d_2\rangle|=0$, $|d_1\rangle$ and $|d_2\rangle$ can be perfectly distinguished. In the intermediate case, i.e., $|\langle d_1|d_2\rangle|=\sin2\theta$, one can only obtain a partial which-way information by means of nonorthogonal quantum state discrimination. 

We utilize the photon's polarization degree of freedom as the WWD. Let the initial state of the WWD be $|h\rangle$. The polarization of the photon in path 1 is rotated due to the interaction between the WWD and the photon. The quantum state after the interaction becomes  $\sqrt{p_1}|0,h\rangle+\sqrt{p_2}|1,s\rangle$, where $|s\rangle=\sin2\theta|h\rangle-\cos2\theta|v\rangle$. After the second balanced BS, the probability of detecting the photon at path 0 is $p=(1+2\sqrt{p_1p_2}\sin2\theta\cos\varphi)/2$. Here $\varphi$ is the phase between the two paths. Thus the visibility is given by
\begin{eqnarray}
V=\frac{p_{\rm max}-p_{\rm min}}{p_{\rm max}+p_{\rm min}}=2\sqrt{p_1p_2}\sin2\theta.
\label{v}
\end{eqnarray}
To retrieve the which-way information, one could perform the UQSD strategy on the polarization of the photon. The maximum probability of unambiguously discriminating the polarization states is given by Eqs.~(\ref{eq4}) and (\ref{eq5}). We now have
\begin{eqnarray}
\label{du1}
D_{\rm u}+V&=&1, \quad p_2/p_1> \sin^22\theta,\\	
D_{\rm u}+V&=&p_1\cos^22\theta+2\sqrt{p_1p_2}\sin2\theta,\quad p_2/p_1\leq \sin^22\theta.
\label{du}
\end{eqnarray}
Equations (\ref{du1}) and (\ref{du}) coincides with the results in Ref.~\cite{menon2018wave} which considers a double-slit scenario. 

On the other hand, the which-way information can also be retrieved by the MED strategy. The probability of the correct guess is given by Eq.~(\ref{pr}). Thus the distinguishability becomes $D_{\rm m}=2P_{\rm r}-1$. Using Eq.~(\ref{v}), we recover
\begin{eqnarray}
D_{\rm m}^2+V^2=1.
\label{qua}
\end{eqnarray}
Since we consider a pure state as the input, the duality relation is an equality. If we consider a more general case (e.g., the input being a mixed state), it would be an inequality. In recent years, constraints of the entanglement on the duality relation Eq.~(\ref{qua}) have been fruitfully discussed in both classical and quantum domains \cite{PhysRevLett.117.153901, qian2018entanglement, DeZela18, quantum2040035, Norrman20, qian2020turning, Ph}. In particular, Ref. \cite{yoon2021quantitative} demonstrated the constraints of the purity of the photon source on the duality relation Eq.~(\ref{qua}) from a source point of view \cite{PhysRevResearch.2.012031} by adjusting the amplitudes of the seed laser, where the photon source was generated through an entangled nonlinear bi-photon source model.

\begin{figure}[tbp!]
\centering\includegraphics[width=0.5\textwidth]{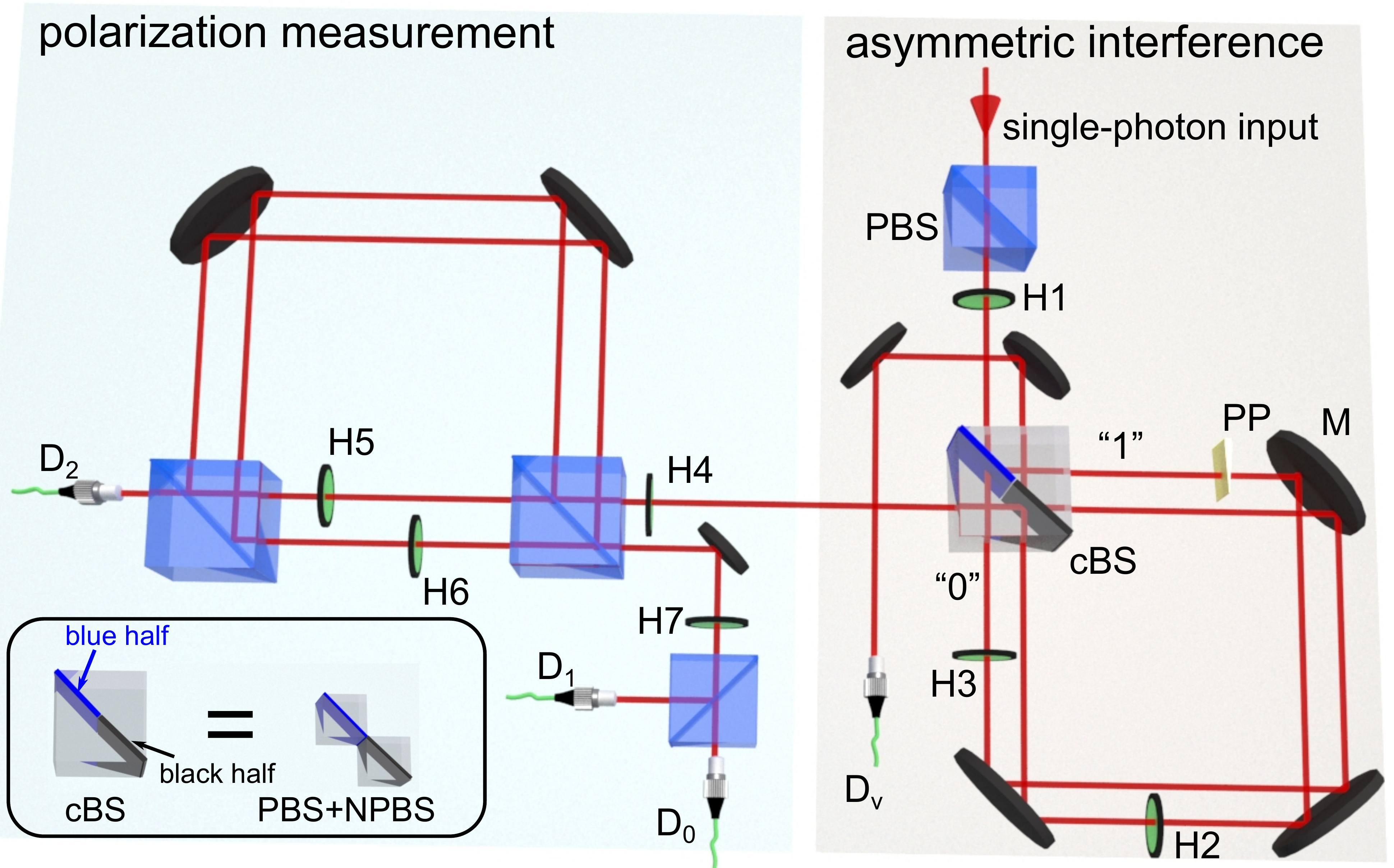}
\caption{Schematic of the experiment. (1): asymmetric interference. The photon transmits through the first PBS and is then rotated by the first half-wave plate H1. The blue half of the cBS works as a PBS to split different polarization components. The photon traverses the two paths in the Sagnac-like structure and recombines on the black half of the cBS. H2 and H3 determine the final states of the WWD. PP introduces a phase between the two paths. (2): polarization measurement. The polarization of the photon is analysed in the Sagnac-like structure. See Methods for the details of the implementation. The inset shows that the cBS is equivalent to an assembly of a PBS and a NPBS. When the photon is incident on the blue region, it works as a PBS; while when the photon is incident on the black region, it works as a NPBS. PBS: polarization beam splitter; NPBS: non-polarizing beam splitter; PP: phase plate; cBS: cubic beam splitter; M: mirror; H1\textasciitilde H7: half-wave plates; D$_{\rm v}$, D$_0$, D$_1$, and D$_2$: single-photon detectors.}
\label{fig3}
\end{figure}
\subsection{Experimental setup}
To implement the forementioned asymmetric beam interference, our experimental setup consists of two Sagnac-like structures. The first Sagnace loop realizes the asymmetric beam interference, while the second one realizes the polarization measurement, as is shown in Fig.~\ref{fig3}. The photon source is a single photon generated through a nonlinear process (See Methods for details). The polarization of the photon is prepared to be horizontal by the first polarization beam splitter (PBS). A half-wave plate (H1) and a cubic beam splitter (cBS) work as a variable beam splitter. The cBS, which is part of the Sagnac-like structure, is customized such that the coating inside the crystal consists of two parts, a blue part and a black part. It functions as a PBS when the photon is incident on the blue half of the crystal, while it functions as a non-polarizing beam splitter when the photon is incident on the black half. The cBS is functionally equivalent to an assembly of a PBS and a non-polarizing beam splitter, as is shown in the inset. Such a compact structure enables relative stability of the Sagnac-like structure. Therefore, when H1 is oriented at $\theta_{\rm a}$ and the photon is incident on the blue half of the cBS, the photon travels along path 0 or path 1, depending on the polarization, with probabilities $p_1=\cos^22\theta_{\rm a}$ and $p_2=\sin^22\theta_{\rm a}$, respectively. The photon in path 0 is vertically polarized while the photon in path 1 is horizontally polarized. The split ratio $p_2/p_1=\tan^22\theta_{\rm a}$ determines the asymmetry of the interference. At this step, since the polarization correlates with the path, the polarization is regarded as the WWD.

Inside the first Sagnac loop, a half-wave plate H2 oriented at $\theta_{\rm n}$ is inserted in path 1 to set the nonorthogonality of the final states of the WWD. To maintain the coherent superposition of the two paths, another half-wave plate (H3) oriented at $0^{\circ}$ is inserted in path 0 to compensate the optical path. The photon from the two paths interferes on the black half of the cBS. The output states, which correspond to the two exits of the cBS, are given by
\begin{eqnarray}
|\psi_{\rm v}\rangle=\frac{1}{\sqrt{2}}(\cos 2\theta_{\rm a}|d_1\rangle-e^{i\varphi}\sin 2\theta_{\rm a}|\overline{d}_2\rangle),\\
|\psi_{\rm d}\rangle=\frac{1}{\sqrt{2}}(\cos 2\theta_{\rm a}|d_1\rangle+e^{i\varphi}\sin 2\theta_{\rm a}|d_2\rangle), 
\label{unnor}
\end{eqnarray}
where $|d_1\rangle=|h\rangle$ and $|d_2\rangle=\sin 2\theta_{\rm n}|h\rangle-\cos 2\theta_{\rm n}|v\rangle$ are the final states of the WWD, whose nonorthogonality is determined by $|\langle d_1|d_2\rangle|=\sin 2\theta_{\rm n}$; and $|\overline{d}_2\rangle=\sin 2\theta_{\rm n}|h\rangle+\cos 2\theta_{\rm n}|v\rangle$. The state $|\psi_{\rm v}\rangle$ is detected immediately by D$_{\rm v}$ for measurement of the visibility
\begin{equation}
V=\frac{{\rm max}(N)-{\rm min}(N)}{{\rm max}(N)+{\rm min}(N)},
\label{vv}
\end{equation}
where $N$ is the photon count at D$_{\rm v}$, ${\rm max(\cdot)}$ and ${\rm min(\cdot)}$ are the extreme values with respect to $\varphi$, which is the phase introduced by the phase plate in path 1. Afterwards, the photon in the state $|\psi_{\rm d}\rangle$ enters the second Sagnac-like structure for the distinguishability measurement. The which-way information of the photon implies the path along which the photon travels in the first Sagnac loop.

\subsection{Experimental linear and quadratic duality relations}
We now perform a nonorthogonal state discrimination on the states $|d_1\rangle$ and $|d_2\rangle$ to measure the distinguishability. We first quantify the distinguishability as the probability of an error-free result, i.e., by adopting the UQSD strategy. The procedure is analogous to discriminating two nonorthogonal states with equal \textit{a priori} probabilities apart from additional basis rotations \cite{clarke2001experimental}. The half-wave plates H4$\sim$H7 are properly rotated to realize the basis transformation (See Methods for details). Here, a click at D$_2$ corresponds to an inconclusive result, indicating that the photon may come from path 0 or path 1. A click at D$_1$ indicates that the photon deterministically comes from path 1, and a click at D$_0$ indicates the photon deterministically comes from path 0. In this setup, the photon count of path 0 is $N_{20}+N_{00}$, and the photon count of path 1 is $N_{21}+N_{11}$, where $N_{ij}$ is the photon count at D$_i$ when path $j$ is open in the first Sagnac loop. The photon count, corresponding to an unambiguous result, is $N_{11}+N_{00}$. The distinguishability $D_{\rm u}$ is quantified by the probability of getting an unambiguous result
\begin{equation}
D_{\rm u}=\frac{N_{00}+N_{11}}{(N_{20}+N_{00})+(N_{21}+N_{11})}.
\label{dv}
\end{equation}

Figure \ref{d2} shows the photon counts at D$_0$, D$_1$, and D$_2$ with respect to $\varphi$ when (a) $\tan2\theta_{\rm a}=0.38$, $\sin2\theta_{\rm n}=0.2$ and (b) $\tan2\theta_{\rm a}=0.28$, $\sin2\theta_{\rm n}=0.9$. The sinusoidal change of the photon count at D$_2$ is due to the interference between the common parts of $|d_1\rangle$ and $|d_2\rangle$. In Fig.~\ref{d2}a, when $\tan2\theta_{\rm a}>\sin2\theta_{\rm n}$, the minimum photon count at D$_2$ is zero. This implies that the common part is equally likely to come from $|d_1\rangle$ or $|d_2\rangle$ (equal lengths of the common parts in Fig.~\ref{fig2}a). While in Fig.~\ref{d2}b, when $\tan2\theta_{\rm a}\leq\sin2\theta_{\rm n}$, the minimum photon count at D$_2$ is zero since the photon in the common part is more likely to come from path 0; in other words, some which-way information is stored in the common part.
\begin{figure}[tbp!]
\caption{Normalized photon counts at D$_0$, D$_1$, and D$_2$. (a) $\tan2\theta_{\rm a}=0.38$, $\sin2\theta_{\rm n}=0.2$ and (b) $\tan2\theta_{\rm a}=0.28$, $\sin2\theta_{\rm n}=0.9$. The photon counts are divided by the total photon count of D$_{\rm v}$+D$_0$+D$_1$+D$_2$. The dots are the experimental data, while the solid lines are the theoretical values. The error of the photon count at D$_1$ in (b) is $\sim\mathcal{O}(10^{-4})$, thus the error bar is almost invisible in the figure. The error bars indicate one standard deviation.}
\label{d2}
\end{figure}

Next we perform the MED strategy to extract the which-way information. As illustrated in Fig.~\ref{fig2}b, to realize the MED strategy, the H4 first rotates the polarization suitably. The H5 and the H7 are constantly kept at $0^{\circ}$ (see Methods for details). Under such configuration, $|d_1\rangle$ and $|d_2\rangle$ are first rotated and then projected onto the $|h\rangle$ and $|v\rangle$ basis. The horizontal component is detected by D$_2$ and the vertical component is detected by D$_0$. When D$_2$ clicks, we guess the photon comes from path 0; while when D$_0$ clicks, we guess the photon comes from path 1. Thus, a right guess, with photon count of $N_{01}+N_{20}$, means either D$_2$ clicks when the photon comes from path 0, or D$_0$ clicks when the photon comes from path 1; while a wrong guess is the opposite, with photon count of $N_{00}+N_{21}$. The distinguishability is quantified by the difference between the right guess and the wrong guess
\begin{eqnarray}
D_{\rm m}=\frac{N_{01}+N_{20}-N_{00}-N_{21}}{N_{01}+N_{20}+N_{00}+N_{21}}.
\end{eqnarray}

\begin{figure}[tbp!]
\centering\includegraphics[width=0.48\textwidth]{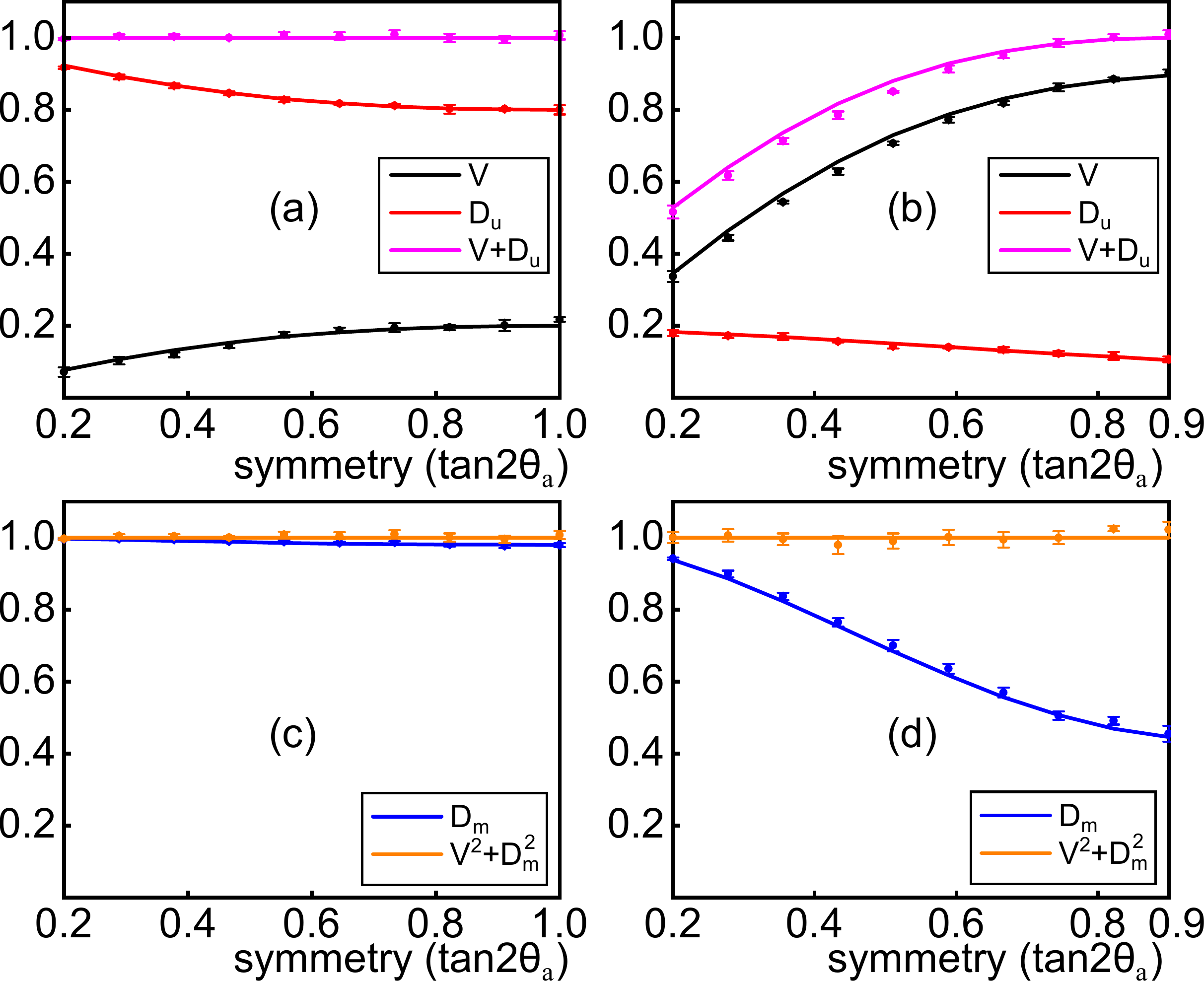}
\caption{Experimental results. Duality relations when the nonorthogonality of the final states of the WWD is (a,c) $\sin2\theta_{\rm n}=0.2$ and (b,d) $\sin2\theta_{\rm n}=0.9$. (a,b): Confirmation of the linear duality relation. (c,d) Confirmation of the quadratic duality relation. The experimental visibility in (c,d) is the same as that in (a,b), both are calculated through Eq.~(\ref{vv}). The label symmetry signifies the symmetry of the interference and it is quantified by $\tan2\theta_{\rm a}$. The dots are experimental values while the solid lines are theoretical values. The error bars indicate one standard deviation.}
\label{fig4}
\end{figure}

Figure \ref{fig4} shows our experimental results, where the horizontal label symmetry signifies the symmetry of the interference and it is quantified by $\tan2\theta_{\rm a}$. We test the two cases when (i) the linear form applies (Fig.~\ref{fig4}a), and (ii) the linear form does not apply (Fig.~\ref{fig4}b), when using the UQSD strategy. For comparison, we also test the quadratic form by using the MED strategy with the same configurations in Fig.~\ref{fig4}c and Fig.~\ref{fig4}d, respectively. Note that the applicability of the linear form requires $\sin2\theta_{\rm n}\leq\tan2\theta_{\rm a}$; therefore, we set $\sin2\theta_{\rm n}=0.2$ in this case to ensure a relatively wide range in which the value of $\tan2\theta_{\rm a}$ could be set. One can see from Fig.~\ref{fig4}a that the experimental summation of $(V+D_{\rm u})$ is close to 1 with small deviations.

On the other hand, for the case where the linear relation does not apply, we set $\sin2\theta_{\rm n}=0.9$ in Fig.~\ref{fig4}b, since the inequality $\sin2\theta_{\rm n}>\tan2\theta_{\rm a}$ should be satisfied. One could see that the linear relation is no longer valid in this case, because partial which-way information is stored in the common part when the degree of asymmetry is larger than the degree of nonorthogonality. On the contrary, in Figs.~\ref{fig4}(c,d), the quadratic relation always applies. One notices that the $(V+D_{\rm u})$ and $(V^2+D_{\rm m}^2)$ in Fig.~\ref{fig4} exceed the theoretical maximum at some data points, this is because the visibility and the distinguishability are not measured with the same photons. The visibility is measured through the statistics of photon counts at detector D$_{\rm v}$ while varying the phase $\varphi$; whereas the distinguishability is measured through the polarization analysis in the second Sagnac loop. On the other hand, the non-ideal coating of the cBS on the black half (i.e., the part acting as an NPBS) causes errors to the measurements of the visibility and the distinguishability. Therefore, the measured $(V+D_{\rm u})$ and $(V^2+D_{\rm m}^2)$ may exceed the theoretical maximum.

\begin{figure}[tbp!]
\centering\includegraphics[width=0.48\textwidth]{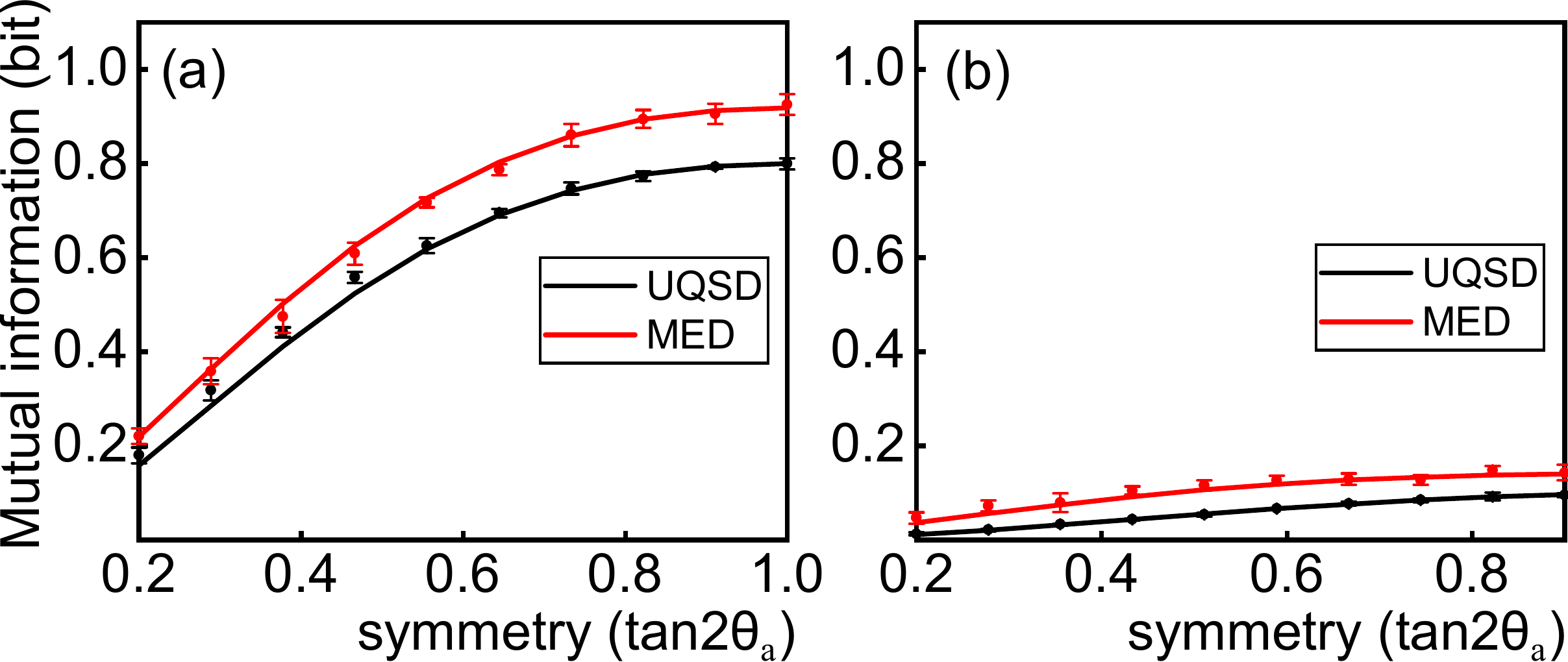}
\caption{Mutual information gained via MED and UQSD strategies. The nonorthogonality is (a) $\sin2\theta_{\rm n}=0.2$ and (b) $\sin2\theta_{\rm n}=0.9$. The dots are experimental values while the solid lines are theoretical values. The error bars indicate one standard deviation.}
\label{mutual}
\end{figure}

\subsection{Relation between the two forms of the duality relation}
The linear duality relation characterized by Eqs.~ (\ref{du1}) and (\ref{du}) is tighter than the quadratic duality relation characterized by Eq.~(\ref{qua}). This can be seen from the mutual information gained after performing the measurement. The mutual information between Alice (A) and Bob (B) is
\begin{eqnarray}
H(A:B)=\sum\limits_{ij}p_i{\rm Tr}(\hat{\rho}_i\hat{\pi}_j)\log\left( \frac{{\rm Tr}(\hat{\rho}_i\hat{\pi}_j)}{{\rm Tr}(\hat{\rho}\hat{\pi}_j)}\right),
\label{mi}
\end{eqnarray}

where the quantum state $\hat{\rho}_i$ is prepared by Alice with \textit{a priori} probability $p_i$, and Bob performs a positive operator-valued measure $\{\hat{\pi}_j\}$ with $\sum_{j}\hat{\pi}_j=\mathbb{I}$ and $\quad\hat{\rho}=\sum_{i}p_i\hat{\rho}_i$. The mutual information given by Eq.~(\ref{mi}) quantifies how much information is obtained by Bob through the measurement (See Methods). The UQSD strategy is closely related to the maximum confidence strategy for quantum state discrimination \cite{mosley2006experimental}, which maximizes the conditional probability $P(\hat{\rho}_i|i)$, i.e., the probability that the state is $\hat{\rho}_i$ when obtaining the result $i$. The MED strategy minimizes the guessing error and in some cases it coincides with the maximum mutual information strategy \cite{levitin1995optimal, barnett2009quantum}. Figure \ref{mutual} shows the mutual information obtained by using USQD and MED strategies. We can see that the mutual information obtained through the MED strategy is more than that obtained through the UQSD strategy. This means that more which-way information is extracted through the MED strategy.

\section{Discussion}
In our work, we measure the visibility by changing the relative phase between the two paths of the first Sagnac loop, while the distinguishability is measured through the polarization measurement in the second Sagnac loop. The visibility and the distinguishability are measured with different photon samples. This, in some sense, implies that they are measured with different setups. While the essence of the duality relations emphasizes the complementarity between the wave behaviour and the particle behaviour of the same photon, we remark that such a method to measure the two quantities has been employed in the study of duality relations \cite{tang2012realization, kaiser2012entanglement}. Due to the destruction of the photon at the detector, we are not able to measure the distinguishability and the visibility with the same photon.

We have realized an asymmetric beam interference experiment to study the wave-particle duality by utilizing the polarization degree of freedom of the photon as a which-way detector. In our experiment, both the linear duality relation and the quadratic duality relation have been confirmed. We have shown that the distinguishability in the linear form corresponds to the probability of obtaining an unambiguous result, while the distinguishability in the quadratic duality relation corresponds to the maximum likelihood for the right guess. We have also shown that the difference between the UQSD strategy and the MED strategy can be understood by calculating the mutual information gained through the measurements. Since less mutual information is gained in the UQSD strategy, the linear form is tighter than the quadratic form. Our results reveal the difference between the two duality relations, which will have fundamental implications in  better understanding the duality relation quantitatively. Furthermore, since the distinguishability is closely related to the discrimination of the states of the which-way detector, our work might motivate future studies on quantum state discrimination in duality relations and may have other potential applications in quantum information science and technology.

\section{Methods}

\textbf{\noindent(A) Details of the experiment.} \\
The single-photon source is generated through spontaneous parametric down-conversion process by pumping a type-I phase matched nonlinear $\beta$-barium-borate crystal. The pump laser is a CW single frequency laser operating at a center wavelength of 404 nm with power of 130 mW. The photon pair with wavelength of 808 nm is filtered by a pair of interference filters with a 3 nm bandwidth. The idler photon is detected by a single-photon detector for coincidence counting. The signal photon is thus heralded and delivered to the experimental setup shown in Fig.~\ref{fig3}. The averaged photon count is approximately 10,000 per second.

The interference visibility of both Sagnac loops is higher than $98.67\%$. The photon counts are measured five times for calculating the deviations, with duration of 0.5 s for each measurement. The error bars in Figs.~\ref{d2}-\ref{mutual} are small because the fluctuation of the photon count is relatively small.

\begin{figure}[tbp!]
\centering\includegraphics[width=0.35\textwidth]{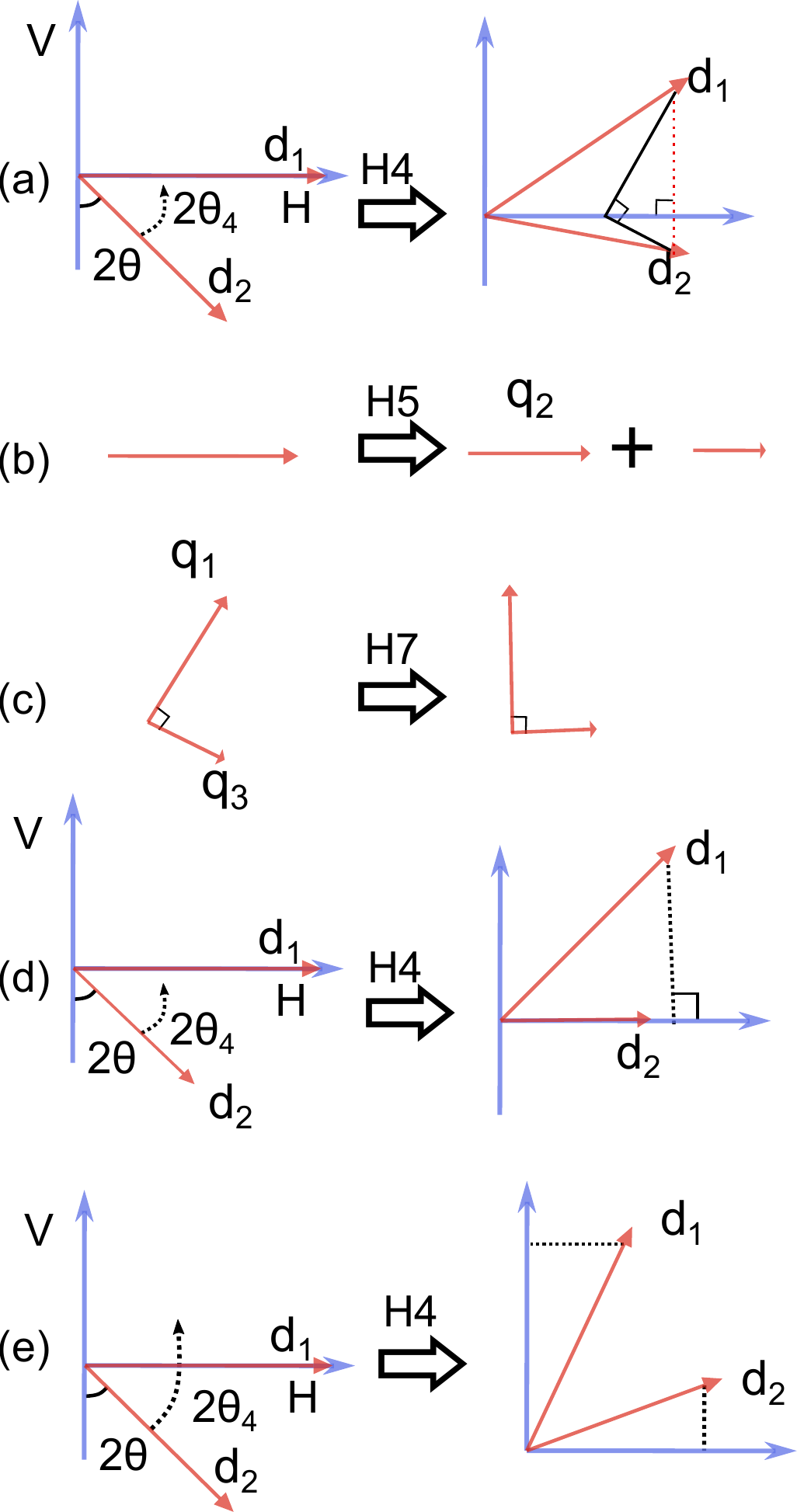}
\caption{Procedures for performing the polarization measurement. (a-d) are the UQSD measurement strategy, (e) is the MED measurement strategy. (a) In UQSD, when $\tan2\theta_{\rm a}>\sin2\theta_{\rm n}$, the H4 rotates the polarization by $2\theta_4$, thereby transforming $|d_2\rangle$ to $|h\rangle$. (b) Then the horizontal component is divided into two parts by H5, one part is the common state ($|d_2\rangle$), and the other is transformed to $|v\rangle$ by H6, which is set at $45^{\circ}$. (c) The residual ($|q_1\rangle$ and $|q_3\rangle$) is unambiguously discriminated by H7 followed by a PBS. (d) When $\tan2\theta_{\rm a}<\sin2\theta_{\rm n}$, the H4 rotates $|d_2\rangle$ to $|h\rangle$. Then the states are projected to the basis states $|h\rangle$ and $|v\rangle$. When we detect the photon in the $|v\rangle$ state, we assert that the photon is in the state $|d_1\rangle$. Otherwise, we obtain an inconclusive result. (e) In MED, the H4 rotates the polarization suitably to minimize the guessing error.}
\label{figs1}
\end{figure}
\vspace{2mm}
\textbf{\noindent(B) Settings for performing UQSD and MED strategies.}\\
We follow Fig.~\ref{fig2} in the main text to clarify the measurement settings for performing the nonorthogonal quantum state discrimination strategies, UQSD and MED.

\noindent(i) UQSD strategy. When $\tan2\theta_{\rm a}>\sin2\theta_{\rm n}$, the H4 rotates the polarization suitably such that the line connecting the endpoints of the state vectors is perpendicular to the horizontal line, thus maximizing the probability of obtaining an unambiguous result. In this way, the states $|d_1\rangle$ and $|d_2\rangle$ have the same amount of horizontal component, as shown in Fig.~\ref{figs1}a. Then H5 separates the horizontal components of both states into two parts. One part corresponds to the common state $|q_2\rangle$ (an inconclusive result); while the other, when superposed with the vertical components, turns the states $|d_1\rangle$ and $|d_2\rangle$ into orthogonal states $|q_1\rangle$ and $|q_3\rangle$. The H6 is fixed at $45^{\circ}$. Finally, $|q_1\rangle$ and $|q_3\rangle$ are unambiguously discriminated by a projective measurement consisting of H7 and a PBS. The angles of H4, H5 and H7 are  
\begin{eqnarray}
\theta_4&=&\frac{1}{2}\arctan\frac{\sin2\theta_{\rm n}-\cot2\theta_{\rm a}}{\cos2\theta_{\rm n}},\\
\theta_5&=&\frac{1}{2}\arccos\frac{\sqrt{\tan2\theta_{\rm a}\sin2\theta_{\rm n}}}{\cos2\theta_4},\\
\theta_7&=&\frac{1}{2}\text{arccot}\frac{\sin2\theta_4}{\cos2\theta_4\sin2\theta_5},
\end{eqnarray}
where $\theta_{\rm a}$ and $\theta_{\rm n}$ are the orientations of H1 and H2, respectively. Here, we omit the reflections on the mirrors. When $\tan2\theta_{\rm a}<\sin2\theta_{\rm n}$, the H4 transforms $|d_2\rangle$ to $|h\rangle$, as is shown in Fig.~\ref{figs1}d. The H5$\sim$H7 are fixed at $0^{\circ}$, $45^{\circ}$ and $0^{\circ}$, respectively. Both of these states have a horizontal component, thus a detection of the $|h\rangle$ state (i.e., D$_2$ clicks) signifies an inconclusive result. While D$_0$ clicks if and only if the state is $|d_1\rangle$. The orientation of H4 is
\begin{eqnarray}
\theta_4=\theta_{\rm n}-\frac{\pi}{4}.
\end{eqnarray}

\noindent(ii) MED strategy. As shown in Fig.~\ref{figs1}e, to realize the MED strategy, the polarization is suitably rotated by H4, followed by a projective measurement. Note that though a simpler setup is sufficient to realize the MED strategy: we maintain the setup unchanged such that it is the same as the one when performing the UQSD strategy. Thus, the H5$\sim$H7 are fixed at $0^{\circ}$, $45^{\circ}$ and $0^{\circ}$, respectively. The orientation of H4 is
\begin{equation}
\theta_4=\frac{1}{4}(\frac{\pi}{2}-\phi),
\end{equation}
where 
\begin{equation}
\phi=\arctan\frac{\cos^22\theta_{\rm a}+\sin^22\theta_{\rm a}\cos4\theta_{\rm n}}{\sin^22\theta_{\rm a}\sin4\theta_{\rm n}}.
\end{equation}

\vspace{2mm}
\textbf{\noindent(C) Evaluation of the mutual information.}\\
We evaluate the mutual information obtained through the polarization measurement for two nonorthogonal states $|d_1\rangle=|h\rangle$ and $|d_2\rangle=\sin 2\theta_{\rm n}|h\rangle-\cos 2\theta_{\rm n}|v\rangle$, with \textit{a priori} probabilities $p_1=\cos^22\theta_{\rm a}$ and $p_2=\sin^22\theta_{\rm a}$, respectively. For UQSD, the states $|d_1\rangle$ and $|d_2\rangle$ are projected onto a three-dimensional space spanned by the basis states $\{|q_1\rangle, |q_2\rangle, |q_3\rangle\}$, with the projective operators $\hat{\pi}_1=|q_1\rangle\langle q_1|$, $\hat{\pi}_2=|q_2\rangle\langle q_2|$, and $\hat{\pi}_3=|q_3\rangle\langle q_3|$. The basis states have the following forms
\begin{eqnarray}
|q_1\rangle&=&R^{\dagger}_{\rm H4}(\hat{P}_v\hat{\sigma_1}\hat{P}_h+\hat{P}_hR^{\dagger}_{\rm H5}\hat{P}_v)R^{\dagger}_{\rm H7}|h\rangle_{r},\\
|q_2\rangle&=&R^{\dagger}_{\rm H4}\hat{P}_hR^{\dagger}_{\rm H5}|h\rangle_{l},\\
|q_3\rangle&=&R^{\dagger}_{\rm H4}(\hat{P}_v\hat{\sigma_1}\hat{P}_h+\hat{P}_hR^{\dagger}_{\rm H5}\hat{P}_v)R^{\dagger}_{\rm H7}|v\rangle_{r},
\end{eqnarray}
where $\hat{\sigma_1}$ is the Pauli operator, $\hat{P}_{h(v)}$ projects the state onto $|h(v)\rangle$, and $R^{\dagger}_{(\bullet)}$ is the Hermitian conjugate of the Jones matrix of the half-wave plate. The subscripts $l$ and $r$ indicate different paths in the three-dimensional space, because the path degree of freedom of the photon is coupled with the polarization degree of freedom to form a higher space. The eigenvalue of $\hat{\pi}_2$ corresponds to an inconclusive result, while the eigenvalues of $\hat{\pi}_1$ and $\hat{\pi}_3$ correspond to unambiguous results that the photon comes from path 0 and path 1, respectively.

While for MED, the projective operators are
\begin{eqnarray}
\hat{\pi}_1&=&R^{\dagger}_{\rm H4}|h\rangle\langle h|R_{\rm H4},\\
\hat{\pi}_2&=&R^{\dagger}_{\rm H4}|v\rangle\langle v|R_{\rm H4},
\end{eqnarray}
which means that the states are projected onto the basis states $|h\rangle$ and $|v\rangle$ after transformed by H4.

\vspace{2mm}
\noindent\textbf{Data availability}\\
All relevant data are available from the corresponding authors upon reasonable request.

\vspace{2mm}
\noindent\textbf{Code availability}\\
All relevant codes are available from the corresponding authors upon reasonable request.

\section*{Acknowledgments}
This work is partly supported by the National Natural Science Foundation of China (NSFC) (Grants No. 11774076, 11804228, U21A20436), Jiangxi Natural Science Foundation (Grant No. 20192ACBL20051, 20212BAB211018 ). F.N. is supported in part by: Nippon Telegraph and Telephone Corporation (NTT) Research, the Japan Science and Technology Agency (JST) [via the Quantum Leap Flagship Program (Q-LEAP), and the Moonshot R\&D Grant Number JPMJMS2061], the Japan Society for the Promotion of Science (JSPS) [via the Grants-in-Aid for Scientific Research (KAKENHI) Grant No. JP20H00134], the Army Research Office (ARO) (Grant No. W911NF-18-1-0358), the Asian Office of Aerospace Research and Development (AOARD) (via Grant No. FA2386-20-1-4069), and the Foundational Questions Institute Fund (FQXi) via Grant No. FQXi-IAF19-06.

\vspace{3mm}
\noindent\textbf{Competing interests}\\
The authors declare no competing interests\\
~\\
\noindent\textbf{Author contributions}\\
D.X.C. and Y.Z. conceived the project. D.X.C. performed the experiment with the help from Y.Z. and J.L.Z. All authors contributed to the numerical results and the writing of the paper. C.P.Y and F.N. supervised the project.

~\\

\end{document}